\documentclass{PoS}

\title{A dynamical mechanism for quark confinement}

\ShortTitle{A dynamical mechanism for quark confinement}

\author{\speaker{Kai Schwenzer}%
        \thanks{I am grateful to Reinhard Alkofer, Christian Fischer and Felipe Llanes-Estrada for helpful discussion. This work was supported by FWF grants M979-N16 and P20592-N16
}\\
       University of Graz \\
       E-mail: \email{kai.schwenzer@uni-graz.at}}

\abstract{Recently there has been progress in the understanding of the confinement mechanism in Landau gauge QCD. The emerging dynamical description in terms of the underlying gauge dependent degrees of freedom goes beyond the static confinement in the quenched limit and has the potential to describe the scale-dependent phenomenon seen in nature, where new hadrons are produced when the system is sufficiently excited. I point out that the confinement mechanism for quarks is rather different from the corresponding one for gluons and that both are embedded in a consistent framework that can describe important qualitative properties of strong interaction physics, like chiral symmetry breaking, spontaneous and anomalous mass generation and a linear rising heavy quark potential with an almost quark-mass independent string tension.}

\FullConference{8th Conference Quark Confinement and the Hadron Spectrum\\
		 September 1-6, 2008\\
		 Mainz. Germany}

\begin{document}

So far the literature on Landau gauge QCD was mostly concerned with its Yang-Mills sector and thereby with the properties of the "glue". In this mechanical analogy these analyses probed its consistency and adhesiveness and it turned out that its characteristic long-range properties are not directly generated by the gluonic dynamics but actually induced by the ghost sector of the theory \cite{Alkofer:2000wg,Alkofer:2008jy}. However, the long-term goal of these studies is surely to understand how this "glue" binds matter and correspondingly to explain the confinement of quarks into hadrons and their properties. \\
An important feature of the strongly interacting vacuum is the spontaneous breaking of chiral symmetry ($\chi$SB) and the resulting large masses of most hadronic states. 
The solution of the Yang-Mills sector, taken as an input, yields dynamically generated constituent quark masses of the right order of magnitude \cite{Fischer:2003rp}. This masses, as well as most hadronic properties, are generated by the dynamics in the scale regime of the order of $\Lambda_{QCD}$ and are rather insensitive to the long-range behavior of the gluonic interaction. This demands detailed numerical computations and to reach results at a quantitative level with functional methods should require to include the vertex functions which feature a pronounced angular dependence \cite{3-gluon}. \\
The other characteristic property of QCD is the absence of colored degrees of freedom in final states. This property is encoded in the infrared (IR) limit of Green's functions and qualitative results can be obtained by a mere IR power counting analysis. The scaling solution with a suppressed gluon and an enhanced ghost \cite{Alkofer:2000wg,Alkofer:2008jy} provides an explanation for gluon confinement via the Kugo-Ojima confinement scenario \cite{Kugo:1979gm} which ensures their absence from the physical state space. However, there is a qualitative difference between gauge degrees of freedom and matter fields since the latter carry a conserved global charge. 
Correspondingly, quarks cannot merely be {\em removed} from the physical spectrum but literally have to be {\em confined} in localized, hadronic states that carry the corresponding charge over. This requires a long-range interaction that permanently binds them which in the idealized quenched limit leads to a linear potential between static sources \cite{Wilson:1974sk}. In contrast in dynamical QCD the interaction is scale dependent and has to describe hadronization in excited systems. Thereby, quark confinement is a rather diverse phenomenon, and a mechanism that only explains the absence of quarks in asymptotic states describes only part of the physics. \\
Recently, a novel mechanism for quark confinement has been proposed that provides such a scale dependent description \cite{Alkofer:2008tt}. In contrast to previous pictures, the mechanism does not simply rely on a divergent gluon propagator that directly confines quarks, but shows a more subtle behavior. Whereas the gluon propagator is actually suppressed, it is the coupling of the gluons to the quarks that overturns the gluonic suppression. In particular, this relies on a kinematic IR singularity of the quark-gluon vertex that is triggered when only the gluon momentum becomes soft and independently of the quark kinematics. This kinematic singularity of the quark-gluon vertex induces a long-range interaction in the four-quark vertex when all external quark momenta are finite but the momentum transfer between them becomes small. In the quenched limit this provides a linear rising potential in coordinate space and thereby an explicit mechanism for the confinement of quarks based on infrared singularities. This IR behavior is confirmed by a numerical solution of the corresponding Dyson-Schwinger equations (DSEs) and the numerical results yield a string tension that is very insensitive to value of the current quark mass \cite{Alkofer:2008tt}. Precisely the strong kinematic divergence of the quark-gluon vertex that induces confinement also provides a mechanism for an anomalous mass generation via the Kogut-Susskind mechanism \cite{Alkofer:2008et}. \\
A recent IR analysis of the fully coupled system of DSEs of dynamical QCD \cite{Schwenzer:2008vt} shows that the quark dynamics does not affect the long-range behavior of the gauge sector at all. This implies that the above mechanical analogy is indeed correct since qualitatively the "glue" is unaffected by the quarks and merely binds them. The IR analysis yields different classes of fixed points in the cases of an infinite, a finite or a vanishing quark mass. In dynamical QCD with finite quark masses the possible fixed points in the quark sector of the theory are reduced compared to the quenched approximation. Whereas in quenched QCD there is in addition to the IR fixed point with a strong kinematic divergence of the quark-gluon vertex also one with a bare vertex, only the latter exists in the unquenched case. Therefore there is no infinite-range interaction between color sources in dynamical QCD but it is screened by virtual quark loops. Strikingly the power counting analysis even provides the screening scale which is precisely of the order of the dynamically generated quark mass and should therefore induce string breaking via quark pairs once the system is sufficiently excited. As shown in fig. 1, even the qualitative form of the heavy quark potential is completely determined by the fixed points of QCD \cite{Schwenzer:2008vt} that are approximately realized over certain scale ranges. \\
There are various recent lattice simulations on rather large lattices that seem to favor an alternative decoupling solution where the gluon becomes massive, cf. \cite{Fischer:2008uz} and refs. therein. As discussed in \cite{Alkofer:2008jy,Schwenzer:2008vt} there are no IR enhanced Green's functions in this case - or in terms of the mechanical analogy: such a glue would not be sticky at all. Correspondingly, in this case the long-range interaction between quarks would have to arise from some independent mechanism without being reflected in the underlying Green's functions. In contrast, the scaling solution discussed above provides a description of many important features of QCD entirely in terms of the latter.

\begin{figure}
\begin{minipage}[c][1\totalheight]{0.32\columnwidth}
\hspace*{0.6cm} \vspace*{-0.9cm} \includegraphics[scale=0.4]{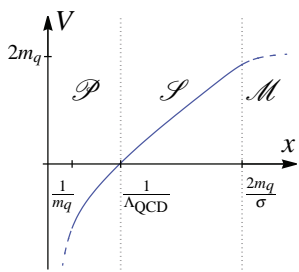}
\flushleft \vspace*{-3.15cm} \hspace*{0.8cm} $V$ \flushleft \vspace*{-0.45cm} \hspace*{0.15cm} $\scriptstyle 2m_q$
\flushleft \vspace*{-0.3cm} \hspace*{0.975cm} ${\mathcal P}$ \hspace*{0.575cm} $\mathcal S$ \hspace*{0.55cm} $\mathcal M$
\flushleft \vspace*{-0.15cm} \hspace*{3.8cm} $x$
\flushleft \vspace*{-0.2cm} \hspace*{0.7cm} $\scriptstyle \frac{1}{m_q}$ \hspace*{0.4cm} $\scriptstyle \frac{1} {\Lambda_{\mathrm{QCD}}}$ \hspace*{0.75cm} $\scriptstyle \frac{2m_q}{\sigma}$ \vspace*{0.3cm}
\end{minipage}
\begin{minipage}[c][1\totalheight]{0.68\columnwidth}
\caption{\label{}Schematic form of the heavy quark potential with areas of application of the different fixed points of Landau gauge QCD \cite{Schwenzer:2008vt}. The short range Coulomb interaction is described by the perturbative UV fixed point ($\mathcal{P}$). The linear rising part with string tension $\sigma$ is determined by the the static IR fixed point ($\mathcal{S}$) that is realized in the limit of an infinite quark mass $m_q\to\infty$. Finally the long ranged screening is described by the massive IR fixed point ($\mathcal{M}$) that represents the IR behavior for finite $m_q$ .}
\end{minipage}
\end{figure}



\begin{thebibliography}{99}

\bibitem{Alkofer:2000wg}
  R.~Alkofer and L.~von Smekal,
  Phys.\ Rept.\  {\bf 353}, 281 (2001)
  [arXiv:hep-ph/0007355].
  
\bibitem{Alkofer:2008jy}
  R.~Alkofer, M.~Q.~Huber and K.~Schwenzer,
  arXiv:0801.2762 [hep-th].
 
\bibitem{Fischer:2003rp}
  C.~S.~Fischer and R.~Alkofer,
  Phys.\ Rev.\  D {\bf 67} (2003) 094020
  [arXiv:hep-ph/0301094].
  
\bibitem{3-gluon}
  R.~Alkofer, M.~Q.~Huber and K.~Schwenzer,
  article in these proceedings.
  
\bibitem{Kugo:1979gm}
T.~Kugo, I.~Ojima, Prog. Theor. Phys. Suppl. \ {\bf 66} (1979) 1.

\bibitem{Wilson:1974sk}
  K.~G.~Wilson,
  Phys.\ Rev.\  D {\bf 10} (1974) 2445.

\bibitem{Alkofer:2008tt}
  R.~Alkofer, C.~S.~Fischer, F.~J.~Llanes-Estrada and K.~Schwenzer,
  arXiv:0804.3042 [hep-ph], Annals Phys. in press;
  R.~Alkofer, C.~S.~Fischer and F.~J.~Llanes-Estrada,
  Mod.\ Phys.\ Lett.\  A {\bf 23} (2008) 1105.

\bibitem{Alkofer:2008et}
  R.~Alkofer, C.~S.~Fischer and R.~Williams,
  arXiv:0804.3478 [hep-ph].

\bibitem{Schwenzer:2008vt}
  K.~Schwenzer,
  arXiv:0811.3608 [hep-ph].
  
\bibitem{Fischer:2008uz}
  C.~S.~Fischer, A.~Maas and J.~M.~Pawlowski,
  arXiv:0810.1987 [hep-ph].

\end{thebibliography}
\end{document}